\documentclass[aps,prl,floatfix,showpacs,amsmath,amssymb,twocolumn]{revtex4}
\usepackage{pslatex}
\usepackage{graphicx}
\usepackage{dcolumn}
\usepackage{bm}
\usepackage{natbib}
\usepackage[usenames]{color}

\begin{document}
\title{Disconnected Glass-Glass Transitions and Diffusion Anomalies in 
a model with two repulsive length scales}

\author{Matthias Sperl}
\affiliation{Institut f\"ur Materialphysik im Weltraum,
Deutsches Zentrum f\"ur Luft- und Raumfahrt, 51170 K\"oln, Germany}
\author{Emanuela Zaccarelli}
\author{Francesco Sciortino}
\affiliation{Dipartimento di Fisica and CNR-INFM-SOFT, Universita` di 
Roma La Sapienza, Piazzale A. Moro 2, 00185 Roma, Italy}
\author{Pradeep Kumar}
\affiliation{Center for Studies in Physics and Biology,
Rockefeller University, New York, NY 10021, USA}
\author{H. Eugene Stanley}
\affiliation{Department of Physics and Center for Polymer Studies, 
Boston University, Boston, Massachusetts 02215, USA}

\date{\today}
\begin{abstract}

Building on mode-coupling-theory calculations, we report a novel scenario 
for multiple glass transitions in a purely repulsive spherical potential: 
the square-shoulder. The liquid-glass transition lines exhibit both 
melting by cooling and melting by compression as well as associated 
diffusion anomalies, similar to the ones observed in water. Differently 
from all previously investigated models, here for small shoulder widths a 
glass-glass line is found that is disconnected from the liquid phase.  
Upon increasing the shoulder width such a glass-glass line merges with the 
liquid-glass transition lines, featuring two distinct endpoint 
singularities that give rise to logarithmic decays in the dynamics. These 
findings can be explained analytically by the interplay of different 
repulsive length scales.

\end{abstract}


\maketitle

In the field of glassy slow dynamics, many experiments and simulations 
have been inspired in recent years by the mode-coupling theory of the 
glass transition (MCT) \cite{Goetze2009}. The theory deals with density 
autocorrelation functions $\phi_q(t)$ with wave vectors $q$, and predicts 
their long-time limits $f_q$. While in the liquid state $f_q=0$, the glass 
state is defined by $f_q > 0$. MCT was first applied to the hard-sphere 
system (HSS) where a liquid-glass transition was identified 
\cite{Bengtzelius1984}, and confirmed by experiments \cite{Megen1991}. In 
addition to liquid-glass transitions, for certain interactions MCT also 
predicts glass-glass transitions: In this case an existing first glass 
state with $f_q^1$ transforms into a second distinct glass state with 
$f_q^2 > f_q^1$ discontinuously. Such glass-glass transitions were 
predicted for the square-well system (SWS) where the hard-core repulsion 
is supplemented by a short-ranged attraction 
\cite{Fabbian1999,Bergenholtz1999,Dawson2001mod}. In the SWS, the first 
glass state is driven by repulsion like in the HSS and the second glass 
state is driven by attraction. The competition between these two 
mechanisms is responsible for the emergence of glass-glass transitions. 
Such a line of glass-glass transitions extends smoothly a line of 
liquid-glass transitions into the glass state and terminates in an 
endpoint singularity. Close to the endpoint singularity the dynamics is 
ruled by logarithmic relaxation \cite{Goetze2002}. The predicted 
logarithmic decays were identified in computer simulations and establish 
the relevance of endpoint singularities for the description of glassy 
dynamics \cite{Sciortino2003,Puertas2003}. A second dynamical anomaly 
predicted for the SWS concerns a reentrant liquid-glass line that causes 
melting by cooling \cite{Fabbian1999,Bergenholtz1999,Dawson2001mod}. This 
prediction of MCT was confirmed by computer simulation 
\cite{Foffi2002bmod} and by experiments in colloidal suspensions 
\cite{Eckert2002,Pham2004}.

\begin{figure}[htb]
\includegraphics[width=.5\columnwidth]{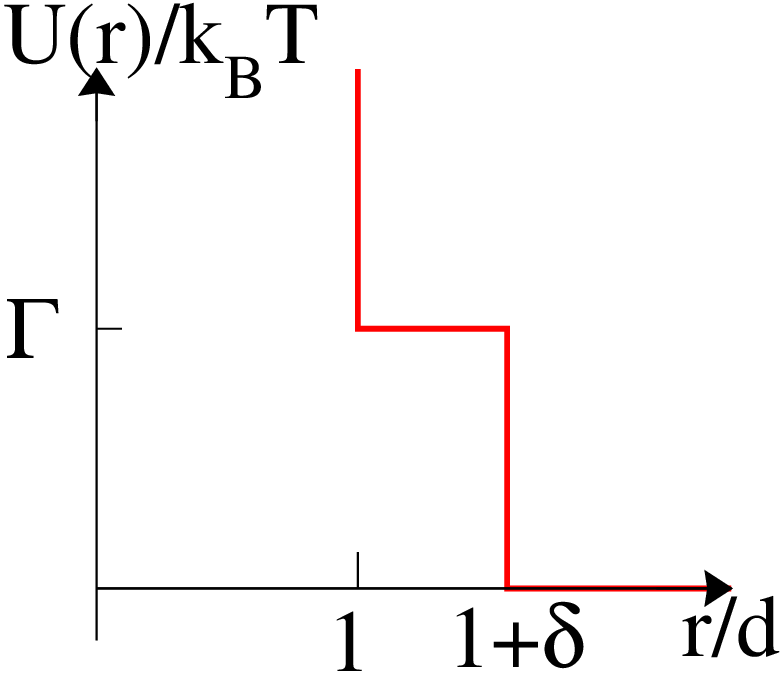}
\caption{\label{fig:sss_pot}Square-shoulder potential with control 
parameters packing fraction $\varphi = \pi\rho d^3/6$, shoulder height 
$\Gamma = u_0/k_\text{B}T$, and shoulder width $\delta$ for particles of 
diameter $d$ at density $\rho$.}
\end{figure}

In this work we replace the attractive length scale in the SWS by a second 
repulsive length scale $\delta$ of the square-shoulder system (SSS) as 
shown in Fig.~\ref{fig:sss_pot}. The SSS can be considered the simplest 
potential with two competing interparticle distances; it is applied to 
describe properties of metallic glasses like cerium or cesium 
\cite{Young1977}, micellar \cite{Osterman2007} and granular materials 
\cite{Duran1999}, silica \cite{Horbach2008} and water \cite{Jagla1999}.

In the following, the glass-transition diagrams are calculated from the 
singularities of the MCT functional \cite{Goetze2009}
\begin{subequations}
\label{eq:FV}
\begin{equation}
\label{eq:F}
{\cal F}_q[\mathbf{V},f_k] =
\sum_{\vec{k}+\vec{p}=\vec{q}}
V_{q,k\,p}\,f_kf_p = 
\frac{f_q}{1-f_q} \,,
\end{equation}
with vertex $\mathbf{V}$ given by wave vector moduli $q$, $k$, $p$, 
the static structure factor $S_q$, and the direct correlation function 
$c_q$:
\begin{equation}
\label{eq:V}
\sum_{\vec{k}+\vec{p}=\vec{q}}
V_{q,k\,p}\,f_kf_p,\;
V_{q,k\,p} =
\rho {S_q S_k S_p}  
\{\vec{q}\cdot[\vec{k}\,{c_k}+\vec{p
}\,{c_p}]\}^2/q^4.
\end{equation}
\end{subequations}
Quantities $S_q$ and $c_q$ are given by the interaction potential. The 
discretization of the functionals is chosen like for the HSS and the SWS 
\cite{Franosch1997,Dawson2001mod} with a number of wave vectors $M = 600$ 
and a wave-vector cutoff $dq_\text{max} = 80$, the static structure 
factors of the SSS are obtained within the Rogers-Young (RY) approximation 
\cite{Rogers1984,Lang1999}; further details of the calculations shall be 
found in a subsequent publication \cite{Sperl2010}. For specific values of 
the control parameters packing fraction $\varphi$, shoulder height 
$\Gamma$, and shoulder width $\delta$, Eq.~(\ref{eq:FV}) exhibits 
singularities where the $f_q$ change discontinuously indicating 
liquid-glass or glass-glass transitions. At these singularities one can 
define the so-called exponent parameter $\lambda<1$ that fixes all 
other critical exponents of the theory \cite{Goetze2009}. While typically 
at liquid-glass transitions, $\lambda\approx 0.7$, $\lambda$ approaches 
unity at the endpoint of glass-glass transition lines signaling the 
emergence of logarithmic decay laws \cite{Goetze2002}.

\begin{figure}[htb]
\includegraphics[width=\columnwidth]{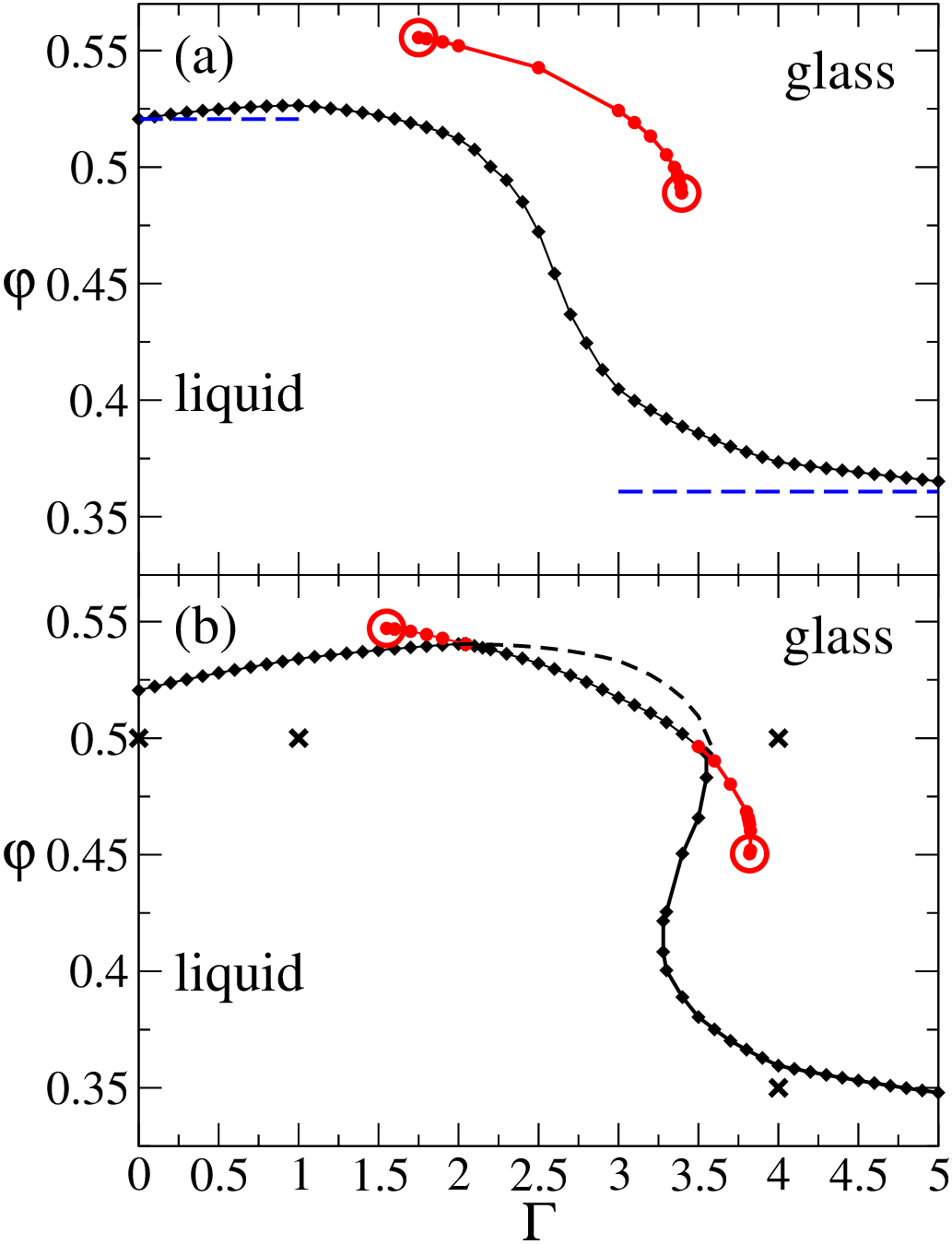}
\caption{\label{fig:PD1513}
Glass-transition diagrams for the SSS. Diamonds ($\blacklozenge$) indicate 
liquid-glass transitions, filled circles ($\color{red}\bullet$) show 
glass-glass transition points terminating in two endpoint singularities 
($\color{red}\circ$). 
(a) $\delta = 0.13$. Dashed lines display the respective limits for hard 
spheres of diameter 1 and $1+\delta$. The line of glass-glass transitions 
is disconnected from the line of liquid-glass transitions, and  moves 
towards and crosses it for larger shoulder width.
(b) $\delta = 0.15$. Crosses ($\times$) indicate the four states 
discussed in Fig.~\ref{fig:beat} relating the glass-glass transitions to 
features in the static structure. The dashed curve exhibits the 
liquid-glass transition line for a reduced wave-vector cutoff 
of $dq_\text{max}=40$ demonstrating that without contributions of higher 
wave vectors to the functional in Eq.~(\ref{eq:FV}) the lines of 
glass-glass transitions are absent.
}
\end{figure}

Figure~\ref{fig:PD1513}(a) shows the glass-transition scenario for $\delta = 
0.13$: The liquid-glass transition line (diamonds) starts at the HSS 
limiting value of $\varphi^c_\text{HSS} = 0.5206$ (dotted line for small 
$\Gamma$) for vanishing shoulder width $\Gamma$. For increasing shoulders 
and up to $\Gamma\approx 1.5$, the curve exhibits a shift in the 
transition packing fraction $\varphi(\Gamma)$ to higher values -- the 
glass initially melts upon cooling. This trend can be traced to the 
evolution of the static structure factor which becomes sharper and moves 
to lower wave vectors for higher $\Gamma$; the corresponding pair 
distribution functions show a higher probability for particles being at 
larger distances from each other. Hence, larger particle separations 
weaken the cage and are compensated by higher densities at the glass 
transition, cf. \cite{Sperl2010} for more details and a comparison to the 
melting-by-cooling phenomenon in the SWS. For shoulder heights from 
$2k_\text{B}T$ to $3.5k_\text{B}T$, i.e. for $2\lesssim \Gamma \lesssim 
3.5$, the glass-transition curve bends downwards and reaches the limiting 
value of the HSS for the outer core $\hat{\varphi}^c_\text{HSS} = 
\varphi^c_\text{HSS}/(1+\delta)^3 = 0.3608$ (dotted line for large 
$\Gamma$).

If the packing fractions and repulsive strengths are increased beyond the 
liquid-glass transition line, one encounters an additional line of 
glass-glass-transition singularities (filled circles). Differently from 
glass-glass lines in all models investigated previously, this additional 
line is located inside the glassy regime, disconnected from any 
liquid-glass transition line. It is bounded by two endpoint singularities 
(open circles) where the additional discontinuity in the $f_q$ vanishes, 
$\lambda$ reaches unity, and logarithmic decay laws emerge.

When increasing the shoulder width $\delta$ further, the glass-glass and 
liquid-glass transition lines move towards each other and start to merge 
for sufficiently high shoulders at around $\delta = 0.145$. 
Figure~\ref{fig:PD1513}(b) shows the situation for $\delta = 0.15$: From 
$\Gamma = 2.0$ to 3.5, the former glass-glass line now indicates a 
transition from the liquid directly into the second glass state. For 
$\Gamma < 2.0$ and for $\Gamma > 3.5$, the formerly isolated glass-glass 
line crosses the liquid-glass line and extends into the glassy regime as 
two glass-glass transition lines. On both ends, the endpoint singularities 
retreat individually in a generic way which is similar to the case of the 
SWS \cite{Dawson2001mod}.

Figure~\ref{fig:PD1513}(b) also exhibits the reentry phenomenon melting-by 
cooling for small $\Gamma$ as discussed before for $\delta = 0.13$, but in 
addition a second reentry phenomenon is found between $\Gamma = 3.0$ and 
3.5 where melting can also be induced by compression. The trends can again 
be traced back to the behavior of the static structure \cite{Sperl2010}: 
Increasing the packing fraction from low values to around $\varphi \approx 
0.45$, the pair distribution function shows an increased probability for 
particle contact at both inner and outer cores. However, for values larger 
than $\varphi \approx 0.45$, the contact at the inner core grows at the 
expense of contact at the outer core; interparticle contacts at the outer 
core are suppressed by the high density and particles now collide much 
more frequently at their inner cores. Since the cage represented by the 
inner core is too loose to trigger glassy arrest at such density, this is 
compensated by lower temperature (or equivalently by higher $\Gamma$) at 
the glass transition.

For the SSS, the interplay between the hard and the soft core is manifest 
in changes at the (1) principal peaks signaling different interparticle 
distances, and (2) by a beating at larger wave vectors that is introduced 
by the oscillation frequencies $q$ and $(1+\delta)q$, respectively. As 
explained above, changes in the principal peaks are sufficient to 
understand the overall behavior of the liquid-glass transition lines, 
including both reentry phenomena.  In the following we demonstrate that 
mechanism (2) is responsible for the glass-glass transitions.

It is known from the theory of Fourier transformations that the 
discontinuities of the direct correlation function in space, $c(r)$, 
determine the large-wave-vector behavior of the direct correlation 
function $c_q$ which enters the MCT functional in Eq.~(\ref{eq:V}) 
\cite{Lighthill1962}. For the square-shoulder potential the dominant 
singularities are inner and outer core which in leading order may be 
modeled by two independent hard-sphere diameters. To obtain a qualitative 
picture, regardless of the specific form of the closure relation, it is 
enough to consider HSS in Percus-Yevick (PY) approximation for which $c_q$ 
is known analytically \cite{Wertheim1963}. For large enough wave vectors
its asymptotic behavior $c_q^\text{asy}$, reads
\begin{equation}\label{eq:cqasy}
c_q^\text{asy} = B(\varphi)\frac{\cos(q)}{q^2} = 
\frac{1+\varphi/2}{(1-\varphi)^2}\frac{4\pi}{q^2}\cos(q)\,.
\end{equation}
$c_1^\text{asy}(q)$ for the inner core and $c_2^\text{asy}(q)$ for the 
outer core are then given by $c_1^\text{asy}(q) = B_1\cos(q)/q^2$ and 
$c_2(q) = B_2\cos(q[1+\delta])/q^2$, where factors of $(1+\delta)$ in 
$q^2$ were absorbed into $B_2$. This analytical PY result for the 
asymptotic decay describes the numerical solution of the RY structure 
factor very accurately as seen in the upper panel of Fig.~\ref{fig:beat}. 
While for the respective limits of small and large hard spheres, 
$c_1^\text{asy}(q)$ and $c_2^\text{asy}(q)$, are relevant only 
individually, for parameters in between, one expects an additive effect 
from both cores. Assuming the PY result for both cores we get for the 
combination $c^\text{asy}(q) = c_1^\text{asy}(q) + c_2^\text{asy}(q)$:

\begin{figure}[bt]
\includegraphics[width=\columnwidth,]{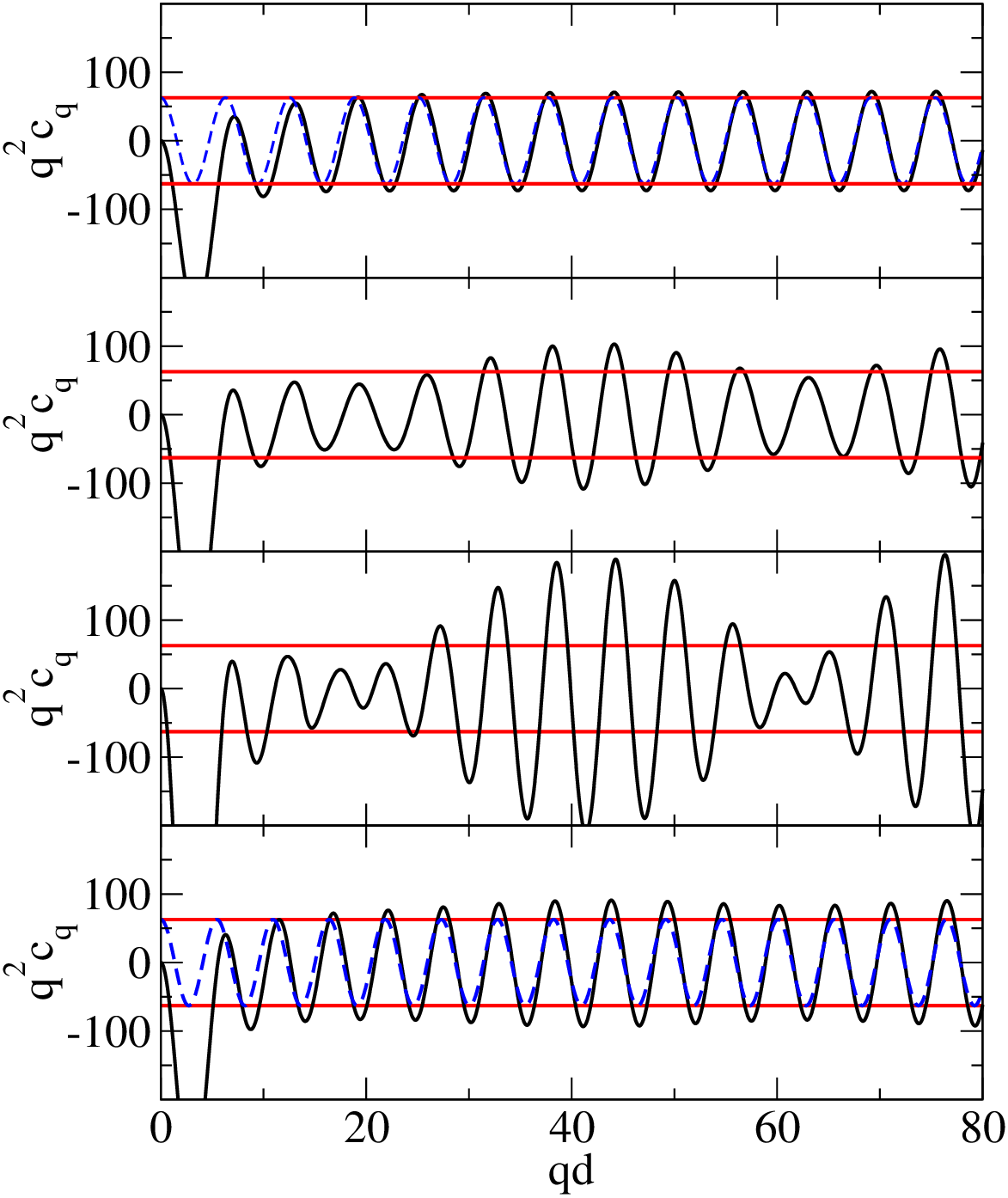}
\caption{\label{fig:beat}
Direct correlation function $q^2c_q$ for $\delta=0.15$, $\varphi=0.50$, 
and $\Gamma = 0.0$, 1.0, and 4.0, respectively from top, $\Gamma 
= 4.0$ and $\varphi=0.35$ in the bottom panel, cf. crosses in 
Fig.~\ref{fig:PD1513}(b). Full horizontal lines display the value of 
$B(\varphi)$ in Eq.~(\ref{eq:cqasy}) for $\varphi=0.50$. The dashed curve 
in the upper panel shows the asymptotic solution of Eq.~(\ref{eq:cqasy}), 
the dashed curve in the bottom panel shows the asymptotic solution of 
Eq.~(\ref{eq:cqasy}) with $q$ replaced by $(1+\delta)q$. The emergence
and vanishing of a beating from top to bottom explains the occurrence of 
the glass-glass transition lines in Figs.~\ref{fig:PD1513}(a) and 
\ref{fig:PD1513}(b) through additional contributions to the functional in 
Eq.~(\ref{eq:FV}).
}
\end{figure}

\begin{equation}\label{eq:beat}
\begin{array}{ll}
c(q)^\text{asy} & = \frac{1}{q^2}\left\{
2B_2\cos\left([1+\delta/2]q\right)
\cos\left(q\delta/2\right)
\right.\\&\left.\qquad+
(B_1-B_2)\cos(q)
\right\}\\& = 
\frac{1}{q^2}\left\{
2B_1\cos\left([1+\delta/2]q\right)
\cos\left(q\delta/2\right)
\right.\\&\left.\qquad+
(B_2-B_1)\cos\left([1+\delta]q\right)
\right\}
\end{array}
\end{equation}
For $B_1 = B_2$, the second lines in Eq.~(\ref{eq:beat}) vanish and a 
simple beating is obtained, for $B_1\neq B_2$ additional terms remain.
Figure~\ref{fig:beat} demonstrates the evolution of the tail in $c_q$ for 
increasing shoulder height. Starting from the HSS of the inner core (upper 
panel of Fig.~\ref{fig:beat}), we get the original $1/q^2$ decay as 
inferred from Eq.~(\ref{eq:cqasy}) or equivalently the second line in 
Eq.~(\ref{eq:beat}). Switching on the shoulder, $B_1> B_2 > 0$, the 
original tail is lowered to $B_1-B_2$ and an additional modulation of 
$\Delta q$ is obtained for the term with mixed frequency. For the 
amplitude variation of the beating one obtains $d\Delta q = 2\pi/\delta$, 
so for $\delta = 0.15$ one gets $\Delta q\approx 42 d^{-1}$, a beating of 
amplitude variation of $qd~\approx 42$ as seen in Fig.~\ref{fig:beat} in 
the two middle panels. If both diameters are of equal importance, the 
beating terms in the first lines of Eq.~(\ref{eq:beat}) become dominant as 
seen in the third panel of Fig.~\ref{fig:beat}. If the outer core is 
dominant, the second equation in Eq.~(\ref{eq:beat}) is relevant: Now the 
decay for the larger core is supplemented by the same beating but with 
prefactor $B_1$. For yet higher contributions from the outer core the 
beating vanishes again as is seen in the bottom panel of 
Fig.~\ref{fig:beat}.

Figure~\ref{fig:beat} explains the origin of the glass-glass transition 
line by the evolution of additional contributions to the MCT vertex in 
Eq.~(\ref{eq:V}). While the upper panel shows a tail in $c_q$ that is 
small enough to no longer influence the MCT vertex, the subsequent panels 
show contributions to the integrals from the beating at wave vectors from 
$qd=20$ to 80 that cannot be neglected. To show that it is these 
contributions that cause the glass-glass transition line we eliminate 
these contributions by shifting the wave-vector cutoff from $qd=80$ to 
$qd=40$ and perform additional calculations. The result is shown in 
Fig.~\ref{fig:PD1513}(b) as the dashed line. With the beating switched off, the 
dashed line of liquid-glass transitions is recovered without any 
indication of crossings or glass-glass transitions. Figure~\ref{fig:beat} 
also demonstrates that the beating is only relevant in a finite region 
that is bound from above and from below which relates to both endpoints of 
the glass-glass transition line.

\begin{figure}[htb]
\includegraphics[width=\columnwidth,]{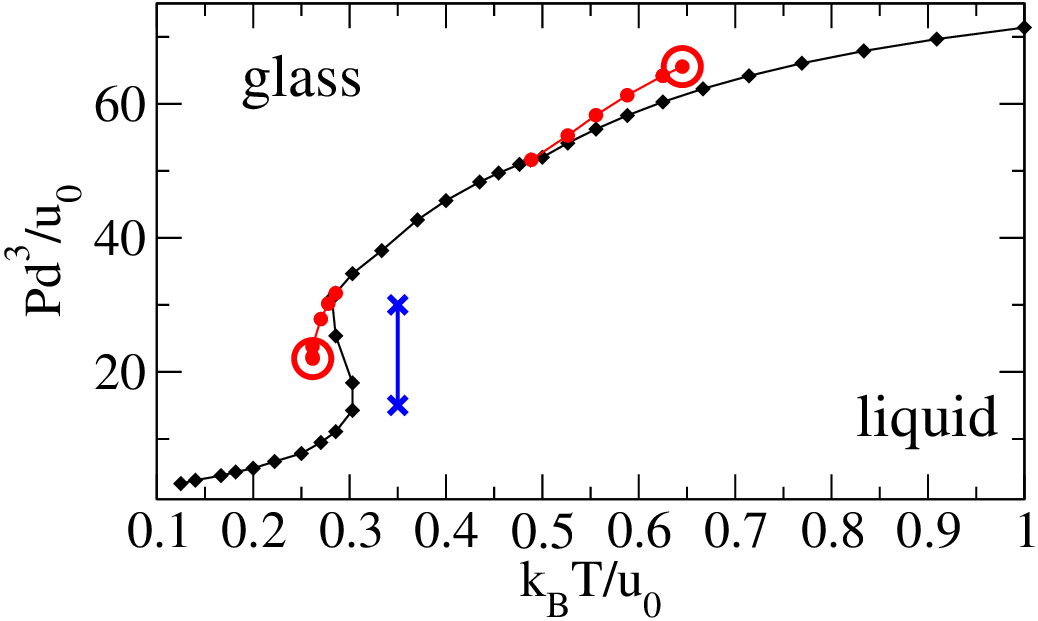}
\caption{\label{fig:PDPT15}Data from Fig.~\ref{fig:PD1513}(b) for $\delta = 
0.15$ shown with variables pressure $P$ and temperature $T$. Crosses 
({\color{blue} $\times$}) delimit the approximate region for a diffusion 
anomaly. Above and below that region, the diffusion coefficient decreases 
with pressure -- within the region the diffusion coefficient increases 
with pressure $P$. 
}
\end{figure}

To make contact with experimental systems, we redraw the transition 
diagram of Fig.~\ref{fig:PD1513}(b) in a pressure-vs-temperature, $P$-$T$, 
diagram in Fig.~\ref{fig:PDPT15}, using the RY thermodynamically 
consistent equation of state. For a path of constant $T$ and variable $P$, 
the diffusivity of the dynamics varies with the distance from the 
liquid-glass transition line. E. g., for $T\approx 0.35 u_0/k_\text{B}$ 
and starting from low $P$, the diffusivity first decreases until $P 
\approx 15u_0/d^3$, then it increases anomalously until around $P \approx 
30u_0/d^3$, and then it decreases again. Such behavior, known as diffusion 
anomaly, is experimentally observed in liquid water \cite{Angell1976}.

In conclusion, MCT calculations predict that the square-shoulder system is 
characterized by an unusual slow dynamics, based on the competition 
between two repulsive length scales.  This manifests itself in an 
unexpected glass-glass transition line which can be completely 
disconnected from the liquid phase. Upon reducing the distance between the 
two repulsive lengths, the glass-glass transition line moves towards the 
glass-liquid line, merging for a special value of the shoulder width and 
giving rise to logarithmic behavior of the decay of correlations in a wide 
region of temperatures and densities. The mechanism for the glass-glass 
transitions is derived analytically from features of the static structure 
factor: The two length scales are reflected in a beating at large wave 
vectors in $S_q$ that cause additional contributions to the MCT functional 
and thus trigger an extra transition.

The peculiar shape of the arrest lines echoes also in the liquid phase, 
generating diffusion anomalies (i.e. non monotonic dependence of the 
diffusion coefficient with the control parameter, either $T$, $\varphi$ or 
$P$).  The relevance of the glass-glass transition scenario found here 
goes beyond the specific SSS potential: The competition between two 
repulsive length scales and the presence of diffusion anomalies are 
observed in several different contexts, ranging from soft matter systems 
\cite{Foffi2003star,Osterman2007} to silica \cite{Horbach2008}, and 
complex liquids \cite{Yan2008,Barraz2009,Gribova2009}. Our work opens a 
perspective for understanding slow dynamics in these disparate systems.

We thank W.~G\"otze, J.~Horbach, A.~Meyer and P.~Ziherl for fruitful 
discussions. MS acknowledges support from DFG Sp714/3-1 and BMWi~50WM0741. 
EZ and FS acknowledge support from ERC-226207-PATCHYCOLLOIDS. PK and 
HES thank NSF Chemistry Division for support.

\bibliographystyle{apsrev}
\bibliography{lit,add}

\end{document}